\documentclass{article}

\usepackage{amsmath}
\usepackage{amssymb}
\usepackage{bm}
\usepackage{comment}
\usepackage{graphicx}
\usepackage{caption}
\usepackage{cite}
\usepackage{upgreek}

\usepackage{booktabs}
\usepackage{array} 
\usepackage{orcidlink}

\newcommand{\cplx}[1]{\bm{#1}}
\newcommand{\mat}[1]{\mathbf{#1}}
\newcommand{\iu}{\mathrm{j}}
\newcommand{\tr}{\mathsf{T}}
\newcommand{\cgamma}{\bm{\mathit{\Gamma}}}
\begin{document}

\title{Design and Repeatability Assessment of an RF MEMS Switching Platform for Automated Microwave SOL Calibration}
\author{Edoardo Boretti$^1$, Nicola~Pompeo$^1$, Enrico~Silva$^1$,\\ Kostiantyn~Torokhtii$^1$, Andrea~Alimenti$^1$
\date{$^1$Department of Industrial, Electronic and Mechanical Engineering (DIIEM), Roma Tre University, 00146 Rome, Italy \\(e-mail: andrea.alimenti@uniroma3.it)}
}



\maketitle

\begin{abstract}
This work presents a compact RF MEMS switching platform for automated one-port SOL calibration of microwave lines. The system is based on a Menlo Micro MM5230 SP4T RF MEMS switch integrated on a grounded coplanar waveguide board and driven by a dedicated MM101-based high-voltage control circuit. The board design is presented, and the repeatability of the calibrated reflection coefficient is assessed through the covariance matrix evaluated from 100 repeated SOL calibration measurements. The platform is tested from 100~kHz to 20~GHz with a Rohde \& Schwarz ZNB43 VNA and the performance compared with an electromechanical microwave switch. The RF MEMS fixture shows lower dispersion of the calibrated reflection coefficient and improved temporal stability, demonstrating its suitability for repeatable automated microwave calibration in measurement environments where compactness, low mass, thermal constraints, and reduced power dissipation are key requirements.
\end{abstract}

\section{Introduction}

Microwave switching is a key function in measurement systems in which multiple signal paths, standards, or devices must be selected without manual reconnection. In many advanced applications, however, the choice of the switching technology is constrained not only by the usual radio-frequency (RF) figures of merit, such as insertion loss, isolation, return loss, and linearity, but also by system-level requirements. Compactness, low mass, reduced power dissipation, and thermal stability become critical in spaceborne payloads, miniaturized microwave systems, cryogenic instruments, and quantum-technology platforms \cite{shao2024rfmemssatellite,spietz2026cryogenic,brewer2025demonstrated,daneshmand2011rf,bardin2021microwaves}. In these contexts, bulky or power-dissipating switching solutions may limit the integration of reconfigurable microwave networks close to the device under test.

This aspect is especially relevant when the measurement reference plane is difficult to place close to the device under test. A common example is represented by cryogenic microwave measurements, where the device under test (DUT) is located inside a cryostat and is connected to the vector network analyzer through long coaxial lines, adapters, and vacuum feedthroughs \cite{booth1994,kitano2008,silva2016,zinsser2019,yeh2013situ,thomas2025cryogenic,arakawa2023calibrated}. Whereas several approximated line calibration schemes have been devised through time by various groups \cite{booth1994,kitano2008,silva2016,zinsser2019}, 
placing a compact switching fixture close to the DUT, and therefore inside the cryogenic environment, is the most direct way to move the calibration reference plane to the actual measurement location and to obtain a high quality calibrated characterization of the DUT \cite{yeh2013situ,thomas2025cryogenic,arakawa2023calibrated}. More generally, whenever the DUT is connected to the instrument through a reconfigurable or state-dependent microwave network, this network cannot be regarded as a passive auxiliary element, but becomes part of the calibrated measurement system. Source-pull and load-pull characterization provide a representative example of this concept, since the DUT is intentionally measured under different impedance conditions and the tuning network must be included in the calibration model \cite{madonna2002testing}.

The development of such fixtures requires a careful comparison between switching technologies. Electromechanical microwave switches are still a demanding reference solution because they are often selected when low insertion loss, high isolation, and high linearity are required \cite{hindle2010state}. Nevertheless, they are macroscopic, mechanically actuated devices. Conventional solid-state switches, on the other hand, offer high speed and long lifetime, but typically at the cost of higher insertion loss and lower linearity \cite{cory2009rf}. RF MEMS switches occupy an intermediate and potentially advantageous position, combining small footprint and low power dissipation with RF performance closer to mechanical switching than to conventional semiconductor switching \cite{shao2024rfmemssatellite,grant2004comparison,kurmendra2024mems,firebaugh2004optical}. In addition to this, recent cryogenic evaluations of commercial SP4T RF MEMS switches have demonstrated their potential for quantum-computing multiplexers, showing stable operation at 5.8~K in the 4--8~GHz qubit band, \cite{lee2026cryogenic,spietz2026cryogenic,spietz2025memsduino}.

A particularly relevant aspect for thermally constrained systems is power dissipation. For an electrostatically actuated MEMS switch, the static power consumption is ideally zero, while the reported dynamic switching energy is about 61~pJ per switching event \cite{lee2026cryogenic}. This corresponds to an average dynamic power of only about 0.61~$\upmu$W for switching rates of 10~kHz. By contrast, conventional current-driven PIN-diode RF switches require a continuous bias current to maintain the selected state; a typical 10~mA bias current with a few-volt bias voltage already corresponds to a static dissipation of the order of several tens of milliwatts per active path \cite{macomMA4AGSW4}. These results suggest that MEMS-based switch networks may become a practical route in cryogenic and space-borne applications. Among the available devices, the Menlo Micro MM5230 \cite{MM5230datasheet} is especially attractive for compact microwave circuits and calibration fixtures. It is a single-pole four-throw RF MEMS switch specified from DC to 18 GHz in normal SP4T operation and up to 26 GHz in Super-Port mode. The device is provided in a 2.5 mm × 2.5 mm BGA package and is specified for low insertion loss, high linearity, high power handling, and a switching lifetime exceeding billions of operations. These features make it a promising candidate for replacing larger electromechanical microwave relays in applications where compactness, low loss, and repeatable switching are simultaneously required.

Motivated by these considerations and by the recent surge of studies on commercial Menlo Micro RF MEMS switches for cryogenic, quantum, and microwave-instrumentation applications \cite{lee2026cryogenic,spietz2026cryogenic,spietz2025memsduino,brewer2025demonstrated,schudak2025glass}, this work investigates an RF MEMS-based switching platform for automated one-port SOL calibration of microwave lines.
The objective of this paper is to evaluate the impact of repeated electrical operations on calibrated measurements. In this work, we present the design and room-temperature experimental assessment of a compact grounded coplanar waveguide switching board based on the MM5230 RF MEMS switch. The board is conceived as a one-port calibration fixture: the common port is connected to the vector network analyzer (VNA), three switched ports are connected to short, open, and load standards, and the fourth port is connected to a generic device under test. By electronically selecting the desired branch, a complete SOL calibration can be performed without a manual reconnection of the standards. The main objective of this work is to evaluate the effect of an RF MEMS switch on the quality of an automated one-port SOL calibration. Although RF MEMS switches show promising standalone specifications, their use in a calibration fixture must be evaluated at the level of the calibrated result. In particular, repeated selection of the short, open, load, and DUT states can introduce state-dependent variations that are not captured by datasheet parameters alone. The repeatability analysis performed here is therefore used to quantify how the switched fixture affects the calibrated DUT reflection coefficient.
The metrological assessment is then performed by repeating the complete SOL calibration sequence through the MEMS switch and evaluating the resulting covariance of the calibrated complex reflection coefficient. The same procedure is repeated using an electromechanical Dow-Key microwave switch, adopted as a reference switching implementation representative of conventional automated RF switching solutions. As supporting contributions, the paper also describes the design and implementation of the complete switching platform, and the control architecture developed to actuate the MEMS switch.

The paper is organized as follows. Section~II describes the hardware design of the proposed platform.  Section~III introduces the metrological background and measurement model. Section~IV reports the experimental results obtained, with emphasis on the calibrated DUT response, the Type-A uncertainty associated with repeated SOL calibrations, and the comparison with an electromechanical Dow-Key switch. Finally, Section~V summarizes the main findings, discusses the present limitations, and outlines possible future developments.

\section{Hardware design}

The designed platform is a compact electronically reconfigurable fixture for automated one-port SOL calibration. Its architecture is based on a single-pole four-throw (SP4T) RF switch, which allows the short, open, load, and DUT states to be selected electronically near the measurement region, thereby moving the effective calibration reference plane from the VNA front panel toward the DUT.

The switching element adopted in this work is the Menlo Micro MM5230 RF MEMS switch \cite{MM5230datasheet}. The device is a micromechanical SP4T switch with independently controlled branches and broadband operation up to 18~GHz in standard SP4T mode.  The main specifications relevant to the proposed fixture are summarized in Table~\ref{tab:mm5230_specs}.

Since the switch is embedded in the calibration fixture,  the RF board was designed with four identical controlled-impedance grounded coplanar waveguide (GCPW) branches routed from the central MEMS switch toward the calibration standards and the DUT port, as described in the following subsection. The associated high-voltage gate-bias and firmware-control architecture is then presented in the subsection \ref{subsec:gate_control}.

\begin{table}[htbp]
\centering
\caption{Main MM5230 specifications relevant to the proposed switching platform.}
\label{tab:mm5230_specs}
\begin{tabular}{lll}
\toprule
Parameter & Typical value / range & Condition \\
\midrule
Topology & SP4T & --- \\
Frequency range & DC--18 GHz & Normal SP4T mode \\
Insertion loss & 0.3 dB/1.3 dB & 6 GHz/18 GHz \\
Return loss & 15 dB/10 dB & 6 GHz/18 GHz  \\
Isolation & 25 dB/18 dB & 6 GHz/18 GHz   \\
IIP3 & 95 dBm & Typical \\
Turn-on time & 8.5 $\upmu$s & Typical \\
Turn-off time & 2.5 $\upmu$s & Typical \\
Gate bias voltage & Typical, $87$--$91$ V range & --- \\
Gate voltage slew rate & 20--200 V/$\upmu$s & --- \\
Switching operations & $>3\times10^9$ & Minimum, 25 $^\circ$C \\
Package size & 2.5 mm $\times$ 2.5 mm & --- \\
\bottomrule
\end{tabular}
\end{table}

\subsection{GCPW Design and Layout}
\label{subsec:gcpw_layout}

The RF board was implemented as a four-layer PCB with the stack-up reported in Table~\ref{tab:pcb_stackup}. The microwave routing is placed on the top copper layer, while the first internal copper layer acts as the main RF reference plane. The two layers are separated by a 0.17-mm-thick Rogers RO4350B laminate. The remaining FR4 layers provide mechanical support and allow the routing of low-frequency control signals without modifying the microwave reference structure.

\begin{table}[htbp]
\centering
\caption{PCB stack-up used for the RF MEMS switching platform.}
\label{tab:pcb_stackup}
\begin{tabular}{lll}
\toprule
Layer & Material & Thickness \\
\midrule
Top copper & Cu & 0.035 mm \\
Dielectric & RO4350B & 0.17 mm \\
Inner copper 1 & Cu & 0.035 mm \\
Dielectric & FR4 & 1.20 mm \\
Inner copper 2 & Cu & 0.035 mm \\
Dielectric & FR4 & 0.20 mm \\
Bottom copper & Cu & 0.035 mm \\
\bottomrule
\end{tabular}
\end{table}

All RF branches were designed as grounded coplanar waveguides (GCPWs). This topology was selected because it provides a planar transmission line with a ground return close to the signal conductor and a lower sensitivity to the lower mechanical layers of the PCB. The first design step was the quasi-static dimensioning of the conductor-backed coplanar waveguide. Following the notation commonly used for this structure \cite{simons2001coplanar}, let $a=w/2$ and $b=a+g$, where $w$ is the signal-conductor width and $g$ is the coplanar gap. The auxiliary parameters are:
\begin{equation}
    k=\frac{a}{b},
    \qquad
    k_3=
    \frac{\tanh\!\left(\pi a/2h\right)}
    {\tanh\!\left(\pi b/2h\right)},
    \label{eq:gcpw_k_parameters}
\end{equation}
defining $k'=(1-k^2)^{1/2}$ and $k_3'=(1-k_3^2)^{1/2}$, the effective permittivity is then evaluated as \cite{simons2001coplanar}:
\begin{equation}
    \varepsilon_{\mathrm{eff}}
    =
    \frac{
    1+\varepsilon_r
    \dfrac{K(k')}{K(k)}
    \dfrac{K(k_3)}{K(k_3')}
    }
    {
    1+
    \dfrac{K(k')}{K(k)}
    \dfrac{K(k_3)}{K(k_3')}
    },
    \label{eq:gcpw_effective_permittivity}
\end{equation}
where $K(\cdot)$ is the complete elliptic integral of the first kind. The corresponding characteristic impedance is
\begin{equation}
    Z_0
    =
    \frac{60\pi}{\sqrt{\varepsilon_{\mathrm{eff}}}}
    \left[
    \frac{K(k)}{K(k')}
    +
    \frac{K(k_3)}{K(k_3')}
    \right]^{-1}.
    \label{eq:gcpw_impedance}
\end{equation}
These equations were used as a first-order design tool to select the line geometry on the RO4350B layer. In the final layout, the signal trace width is $w=0.250$~mm and the coplanar gap is $g=0.100$~mm. These dimensions were kept constant along the RF paths, from the coaxial connectors to the MM5230 switch, without introducing local tapers. This was possible because the selected trace width is compatible with the RF landing geometry of the MEMS package, thus avoiding an additional transition close to the switch.

The RF ports use WithWave NE06FS001 solderless edge-launch SMA connectors. A solderless transition was selected to reduce assembly-dependent variations at the connector-to-GCPW interface. In soldered edge-launch connectors, the amount and shape of solder can introduce small but poorly reproducible changes in the local transition geometry \cite{vasjanov2024unveiling}. This aspect is critical in the present fixture, where the four switched RF paths must be as geometrically equivalent as possible: uncontrolled differences among the short, open, load, and DUT transitions would be directly included in the raw reflection coefficients used for the SOL calibration.

\begin{figure}
    \centering
    \includegraphics[width=1\linewidth]{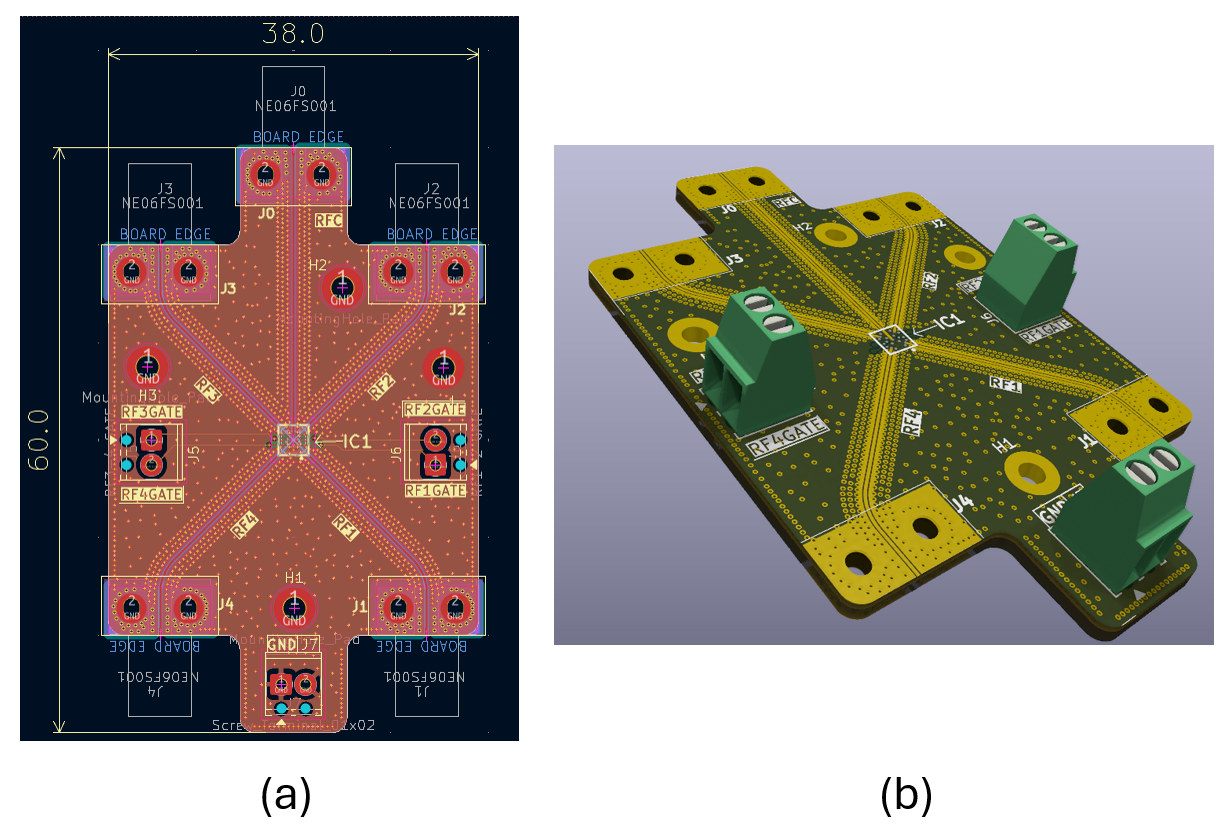}
    \caption{(a) PCB layout of the GCPW board designed to test the MM5230. 
(b) Three-dimensional view of the assembled board.}
\label{fig:RFboardPIC.png}
\end{figure}

The MM5230 switch is placed close to the center of the RF board as shown \figurename~\ref{fig:RFboardPIC.png}. The common branch connects the VNA port to the switch input, while the four output branches are routed toward the short, open, load, and DUT ports. 

Ground continuity is reinforced by a via fence along the GCPW paths. The vias have a diameter of 0.2~mm and a pitch of 0.5~mm, connecting the top coplanar ground to the internal and bottom ground planes. This via stitching improves the equipotentiality of the ground regions, confines the electromagnetic field around the signal conductors, and reduces the risk of exciting parasitic propagation modes \cite{sain2016impact}.

\begin{figure}[htbp]
    \centering
    \includegraphics[width=\columnwidth]{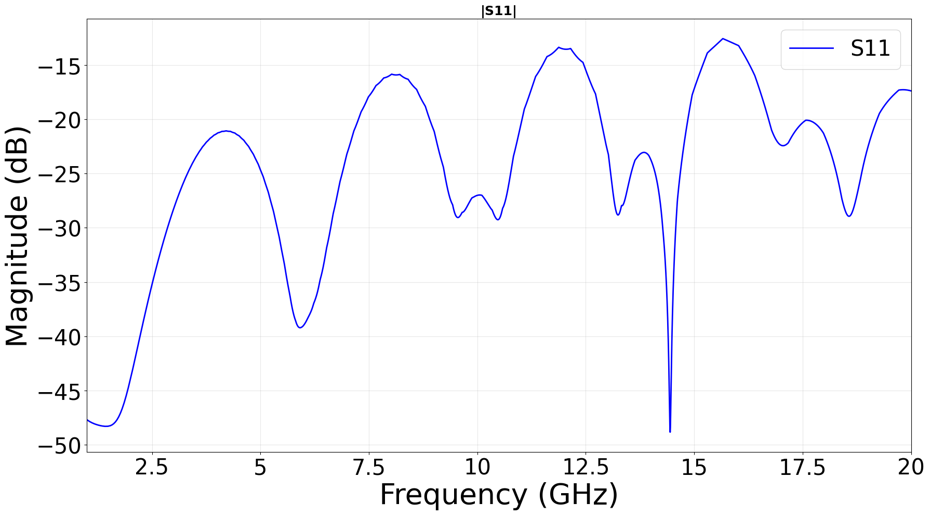}
    \caption{Representative CST simulation of the input reflection coefficient of two concatenated RF paths of the GCPW switching fixture.}
    \label{fig:cst_s11}
\end{figure}

The layout was finally verified by full-wave electromagnetic simulations in CST Studio Suite. The KiCad layout was imported into the simulator and representative RF paths were analyzed over the 1--20~GHz band, including the GCPW routing and the local discontinuities associated with the switch and connector regions. The simulations were used to check the broadband matching of the RF branches and to identify possible layout-induced resonances before fabrication. A representative simulated reflection coefficient is shown in Fig.~\ref{fig:cst_s11}. The obtained response is always below $-12$~dB, confirming the suitability of the layout for broadband one-port calibration experiments. Residual mismatch is expected from the unavoidable discontinuities introduced by the compact switch footprint and the connector launches; these effects are included in the raw measurements and are therefore handled by the SOL calibration procedure.

The board outline is 38~mm $\times$ 60~mm. The RF section is kept compact and separated from the low-frequency control routing as much as possible. Gate-bias and auxiliary signals are routed away from the GCPW regions, whereas the microwave paths remain confined to the top layer over the RO4350B dielectric.

\subsection{Gate-Bias Driver and Firmware Control}
\label{subsec:gate_control}

The MM5230 RF MEMS switch requires high-voltage gate signals to actuate the selected RF branch. For this reason, the RF board was interfaced to a dedicated low-frequency control board based on the Menlo Micro MM101 high-voltage driver \cite{MM101datasheet}. The MM101 is an eight-channel high-voltage parallel driver with push-pull outputs and an integrated charge pump. Starting from a 5~V supply, the device generates the high-voltage rail required by the MEMS gates and drives the selected HV outputs under digital control.

The MM101 avoids the need for an external high-voltage supply and provides an integrated solution for gate-bias generation and distribution. In the present implementation, the device is operated in SPI mode and is controlled by an STMicroelectronics STM32 microcontroller through a firmware layer exposing a USB interface.
The 90~V setting was selected to match the recommended actuation voltage range of the RF MEMS switch. Since only one RF branch is selected at a time, the firmware maps the short, open, load, and DUT states into mutually exclusive MM101 output configurations. This prevents simultaneous actuation of multiple MEMS paths and makes the microwave measurement sequence reproducible.

\begin{figure}
    \centering
    \includegraphics[width=1\linewidth]{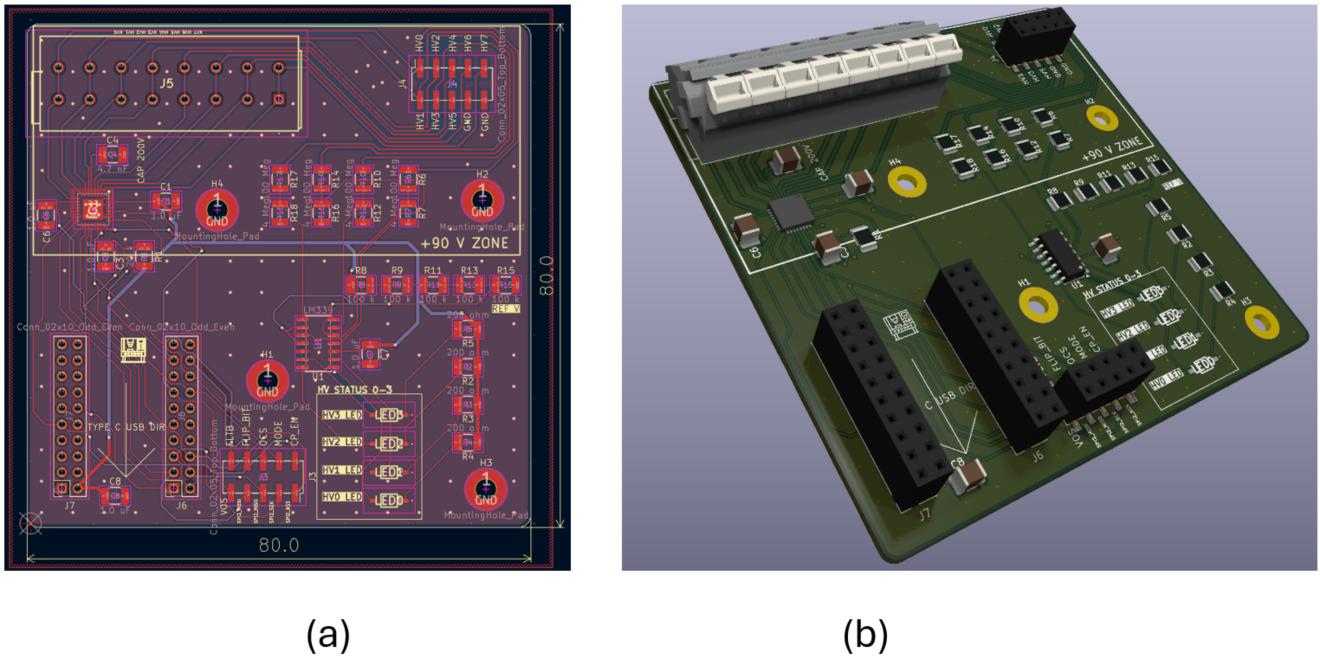}
    \caption{(a) PCB layout of the MM101 board designed for the automated control of the MM5230 switch. 
(b) Three-dimensional view of the assembled board.}
    \label{fig:mm10board}
\end{figure}

The control PCB was implemented as a two-layer FR4 board with a nominal thickness of 1.6~mm and an outline of 80~mm $\times$ 80~mm. It hosts the MM101 driver, the charge-pump and supply decoupling capacitors, high-value biasing and sensing resistors, diagnostic LEDs, connectors for the microcontroller interface, and terminal blocks for the high-voltage gate outputs. The board layout and three-dimensional view are shown in \figurename~\ref{fig:mm10board}.


A critical practical aspect of the control board is the very high impedance required by the MM101 high-voltage outputs. The datasheet specifies a minimum load resistance of 40~M$\Omega$ for operation with the internal charge pump and recommends measurement probes with an input resistance of at least 100~M$\Omega$ to avoid VPP drop. This constraint makes the layout and assembly cleanliness particularly important: flux residues, dust, humidity, or contamination between the closely spaced MM101 pins and the high-voltage traces can introduce leakage paths large enough to prevent the charge pump from reaching the required actuation voltage. For this reason, careful PCB cleaning and inspection were necessary before operation, and the high-voltage region was treated as a sensitive high-impedance analog section.

\section{Metrological Background and Measurement Model}
\label{sec:metrological_model}

In this section, the one-port calibration problem is recalled and the uncertainty terms considered in this work are defined according to a measurement-model approach.

\subsection{One-port VNA calibration problem}
\label{subsec:one_port_calibration}
Let $\cgamma$ be the complex reflection coefficient of the device connected to the measurement reference plane and let $\cgamma_m$ be the raw reflection coefficient indicated by the VNA before error correction. In a conventional one-port model, the relation between $\cgamma_m$ and $\cgamma$ is written as \cite{dunsmore2020handbook}:
\begin{equation}
    \cgamma_m =
    \cplx{E}_{00} +
    \frac{\cplx{E}_{01}\cgamma}
    {1-\cplx{E}_{11}\cgamma},
    \label{eq:one_port_error_model}
\end{equation}
where $\cplx{E}_{00}$, $\cplx{E}_{01}$, and $\cplx{E}_{11}$ are the complex one-port error coefficients, commonly associated with directivity, reflection tracking, and source match, respectively. A short--open--load (SOL) calibration determines these three coefficients by successively measuring three known standards, namely a short, an open, and a matched load. In the ideal case, their reflection coefficients are $\cgamma_S=-1$, $\cgamma_O=+1$, and $\cgamma_L=0$, respectively. Real standards, however, are not ideal lumped terminations: they are calibration-kit artifacts with frequency-dependent complex responses, denoted here as $\cgamma_S(f)$, $\cgamma_O(f)$, and $\cgamma_L(f)$. Their values are obtained from the standard definitions associated with the calibration kit, either in data-based form or through polynomial approximations of their frequency response.

After the error coefficients have been estimated, the reflection coefficient of an unknown DUT is obtained by inverting \eqref{eq:one_port_error_model}:
\begin{equation}
    \hat{\cgamma} =
    \frac{\cgamma_m-\hat{\cplx{E}}_{00}}
    {\hat{\cplx{E}}_{11}
    \left(\cgamma_m-\hat{\cplx{E}}_{00}\right)
    +\hat{\cplx{E}}_{01}},
    \label{eq:error_correction}
\end{equation}
where the hat denotes an estimate. In the present work, the SOL standards are permanently connected to three branches of an SP4T RF MEMS switch, while the fourth branch is connected to the DUT. The calibration sequence is therefore performed by electronically selecting the short, open, and load states instead of manually reconnecting the standards.

For each frequency point $f_k$, the complete switched-calibration procedure can be regarded as the mapping
$\hat{\cgamma}_{\mathrm{DUT}}(f_k)
    =
    g\!\left[\mat{\eta}(f_k)\right],
    \label{eq:measurement_model_general}
$
where $\mat{\eta}(f_k)$ collects the raw complex reflection-coefficient measurements acquired when the short, open, load, and DUT states are selected, namely $\cgamma_{m,S}(f_k)$, $\cgamma_{m,O}(f_k)$, $\cgamma_{m,L}(f_k)$, and $\cgamma_{m,D}(f_k)$, together with the assigned complex values of the SOL standards, $\cgamma_S(f_k)$, $\cgamma_O(f_k)$, and $\cgamma_L(f_k)$. 

For the purposes of the present work, this model is sufficient to isolate the repeatability contribution of the switched fixture. A complete uncertainty evaluation of the absolute DUT reflection coefficient would require a more complete model \cite{zeier2018guidelines}. 

\subsection{Uncertainty Evaluation for Complex Reflection Coefficients}
\label{subsec:uncertainty_complex}
By default, VNAs return the measured reflection coefficient in polar form, namely magnitude in dB and phase in degrees. However, in agreement with the GUM approach for multivariate quantities \cite{jcgm102_2011} and with the recommendations in \cite{hall2016evaluating}, the measured data are immediately converted into complex Cartesian form before performing the uncertainty analysis. This choice avoids the ambiguities associated with phase wrapping and the nonlinear behavior of magnitude and phase, especially when the reflection coefficient is close to the origin of the complex plane.

For the $i$-th repetition, the raw complex reflection coefficient measured in the switch state $q$ is:
\begin{equation}
    \cgamma_{m,q}^{(i)}(f_k)
    =
    10^{M_q^{(i)}(f_k)/20}
    \exp\!\left[
    \iu\frac{\pi}{180}\Phi_q^{(i)}(f_k)
    \right],
    \label{eq:db_phase_to_complex}
\end{equation}
where $q\in\{S,O,L,D\}$ denotes the short, open, load, and DUT states, respectively, $M_q^{(i)}$ is the measured magnitude in dB and $\Phi_q^{(i)}$ is the measured phase in degrees. Each complex reflection coefficient can therefore be expressed as
 $
    \cgamma_{m,q}^{(i)}(f_k)
    =
    \mathit\Gamma_{m,q,\mathrm{re}}^{(i)}(f_k)
    +
    \iu
    \mathit\Gamma_{m,q,\mathrm{im}}^{(i)}(f_k)
$,    
where $\mathit\Gamma_{m,q,\mathrm{re}}^{(i)}$ and $\mathit\Gamma_{m,q,\mathrm{im}}^{(i)}$ are real-valued quantities.

For each repetition $i$, an independent SOL calibration is computed from the three raw measurements $\cgamma_{m,S}^{(i)}(f_k)$, $\cgamma_{m,O}^{(i)}(f_k)$, and $\cgamma_{m,L}^{(i)}(f_k)$. This gives a set of complex one-port error coefficients $\hat{\cplx{E}}_{00}^{(i)}(f_k)$, $\hat{\cplx{E}}_{01}^{(i)}(f_k)$, and $\hat{\cplx{E}}_{11}^{(i)}(f_k)$. The DUT measurement acquired during the same repetition is then error-corrected according to \eqref{eq:error_correction} as:
\begin{equation}
    \hat{\cgamma}_{D}^{(i)}(f_k)
    =
    \frac{
    \cgamma_{m,D}^{(i)}(f_k)-\hat{\cplx{E}}_{00}^{(i)}(f_k)
    }{
    \hat{\cplx{E}}_{11}^{(i)}(f_k)
    \left[
    \cgamma_{m,D}^{(i)}(f_k)-\hat{\cplx{E}}_{00}^{(i)}(f_k)
    \right]
    +
    \hat{\cplx{E}}_{01}^{(i)}(f_k)
    } .
    \label{eq:dut_error_correction_repeated}
\end{equation}
The complete repeated-calibration experiment therefore provides, at each frequency point, a population of $N=100$ calibrated complex reflection coefficients,
$\hat{\cgamma}_{D}^{(1)}(f_k),\ldots,\hat{\cgamma}_{D}^{(N)}(f_k)$.
The mean calibrated reflection coefficient is evaluated as the complex sample mean:
\begin{equation}
    \overline{\cgamma}_{D}(f_k)
    =
    \frac{1}{N}
    \sum_{i=1}^{N}
    \hat{\cgamma}_{D}^{(i)}(f_k).
    \label{eq:complex_mean_dut}
\end{equation}
For the Type-A uncertainty evaluation, each calibrated complex value is represented by the two-dimensional real vector:
\begin{equation}
    \mat{\xi}_i(f_k)
    =
    \begin{bmatrix}
        \hat{\mathit\Gamma}_{D,\mathrm{re}}^{(i)}(f_k) \\
        \hat{\mathit\Gamma}_{D,\mathrm{im}}^{(i)}(f_k)
    \end{bmatrix},
    \qquad
    \overline{\mat{\xi}}(f_k)
    =
    \begin{bmatrix}
        \overline{\mathit\Gamma}_{D,\mathrm{re}}(f_k) \\
        \overline{\mathit\Gamma}_{D,\mathrm{im}}(f_k)
    \end{bmatrix}.
    \label{eq:Cartesian_vector}
\end{equation}
The experimental covariance matrix of the repeated calibrated measurements is then:
\begin{equation}
    \mat{v}_{A}(f_k)
    =
    \frac{1}{N-1}
    \sum_{i=1}^{N}
    \left[
    \mat{\xi}_i(f_k)-\overline{\mat{\xi}}(f_k)
    \right]
    \left[
    \mat{\xi}_i(f_k)-\overline{\mat{\xi}}(f_k)
    \right]^{\tr}.
    \label{eq:typeA_cov_single}
\end{equation}
Equation~\eqref{eq:typeA_cov_single} describes the repeatability of one complete switched-SOL calibration followed by DUT error correction. It is therefore the relevant quantity for assessing the dispersion introduced by the automated calibration procedure.

Since the reported DUT response is obtained as the average of the $N$ calibrated measurements, the Type-A covariance matrix associated with the mean is obtained as
$\mat{U}_{A}(\overline{\cgamma}_{D})(f_k)=\mat{v}_{A}(f_k)/N$.

Although the uncertainty analysis is carried out in Cartesian coordinates, scalar uncertainty indicators in magnitude and phase are useful for comparison with standard VNA outputs. These quantities are obtained by propagating the Cartesian covariance matrix. Denoting by $x$ and $y$ the real and imaginary components of $\overline{\cgamma}_{D}$, the corresponding polar coordinates are $\rho=|z|$ and $\phi=\arg(z)$. i.e. the modulus and phase of the complex quantity $z=x+\mathrm{j}y$. The sensitivity (Jacobian) matrix $\mat{C}_{\mathrm{pol}}$ of the transformation from Cartesian to polar coordinates is:
\begin{equation}
    \mat{C}_{\mathrm{pol}}
    =
    \begin{bmatrix}
        \frac{x}{\rho} & \frac{y}{\rho}
        \\
        -\frac{y}{\rho^2} & \frac{x}{\rho^2}
    \end{bmatrix}.
    \label{eq:jacobian_cart_to_pol}
\end{equation}
The covariance matrix associated with magnitude and phase is therefore:
\begin{equation}
    \mat{U}_{\rho\phi}(f_k)
    =
    \mat{C}_{\mathrm{pol}}
    \mat{U}_{xy}(f_k)
    \mat{C}_{\mathrm{pol}}^{\tr},
    \label{eq:polar_covariance}
\end{equation}
where $\mat{U}_{xy}(f_k)$ denotes the Cartesian covariance matrix to be propagated. In particular, $\mat{U}_{xy}=\mat{v}_A$ when the repeatability of a calibration cycle is considered, whereas $\mat{U}_{xy}=\mat{U}_{A}(\overline{\cgamma}_{D})$ when the uncertainty of the averaged DUT response is reported.

\begin{figure*}[ht]
    \centering
    \includegraphics[width=1\textwidth]{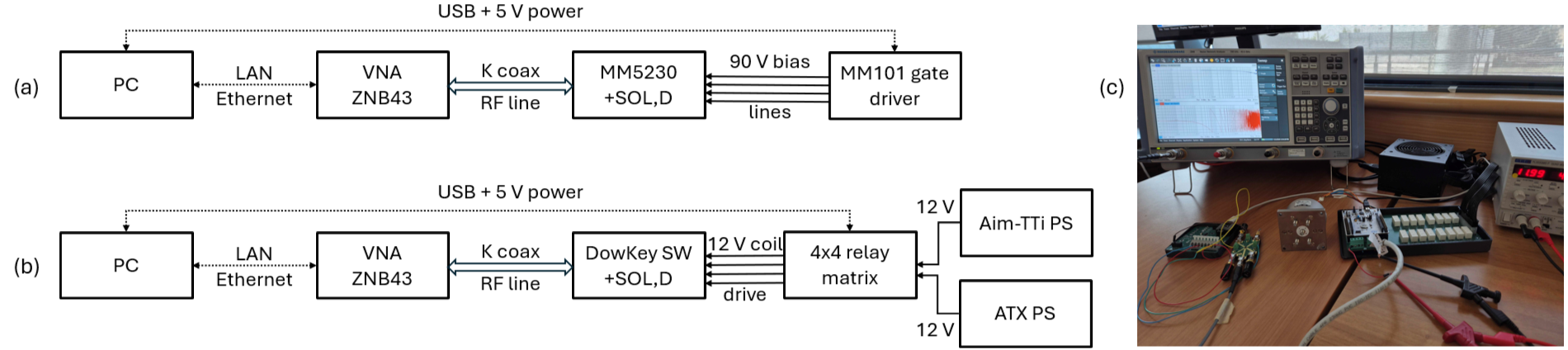}
    \caption{Block diagrams of the measurement setups used for the repeatability analysis: (a) MM5230 RF MEMS-based switching platform and (b) Dow-Key electromechanical-switch setup. In (c) a picture of the systems.}
    \label{fig:measurement_setup_block_diagram}
\end{figure*}

\section{Experimental Section}
\label{sec:results}

The experimental validation was carried out with a Rohde \& Schwarz ZNB43 vector network analyzer. The VNA was configured for one-port S-parameter acquisition over the frequency range from 100~kHz to 20~GHz, using 5000 logarithmically spaced frequency points, an IF bandwidth of 1~kHz, and a source power of 0~dBm. For each repetition, the short, open, load, and DUT states were sequentially selected, with a settling time of 0.5~s after each switching command before triggering the VNA sweep. A total of $N=100$ complete SOL calibration cycles was acquired. The SOL standards, and their characterized frequency behaviour $\cgamma_S(f_k)$, $\cgamma_O(f_k)$, $\cgamma_L(f_k)$, were taken from an Anritsu 3652 calibration kit.

In the MEMS-based setup, the four states were selected by the MM101 gate-driver board through mutually exclusive high-voltage configurations. After each switching command, the firmware applied the cached MM101 state, verified the driver registers and fault status, and only then allowed the VNA acquisition to proceed.
For the electromechanical-switch comparison, the same SOL/DUT acquisition sequence was repeated using a Dow-Key 541-520803 SP4T SMA microwave switch. In this case, the switch actuation was performed by an Aim-TTi power supply routed to the switch coils through the relay switching platform described in \cite{boretti2026relaymatrix}.  The two measurement configurations used for the repeatability comparison are summarized in Fig.~\ref{fig:measurement_setup_block_diagram}.

The DUT used in this experiment was a nominal 57~$\Omega$ through-hole resistor. This component was selected only as a practical test load and not as a calibrated impedance standard. Its absolute impedance cannot therefore be used as a reference value, nor can it be assumed to remain exactly unchanged after successive handling. This means that the DUT was kept mounted for all measurements, and the focus was on repeatability of the effect of the calibration.

\begin{figure}[tbp]
    \centering
    \includegraphics[width=\columnwidth]{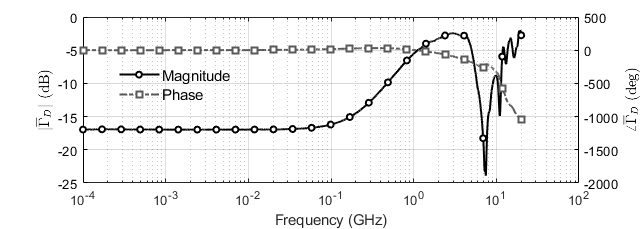}
    \caption{Mean calibrated reflection coefficient of the DUT measured with the MEMS-switched calibration fixture: magnitude and phase.}
    \label{fig:gamma_dut_mems}
\end{figure}

Figure~\ref{fig:gamma_dut_mems} shows the mean calibrated reflection coefficient of the DUT obtained from the MEMS-switched measurements. The magnitude is approximately constant at low frequency and then departs from the purely resistive behavior above the MHz--GHz transition region. The phase response also exhibits a smooth frequency dependence, consistent with a predominantly reactive response of the sample at the higher frequencies. 

\begin{figure}[htbp]
    \centering
    \includegraphics[width=\columnwidth]{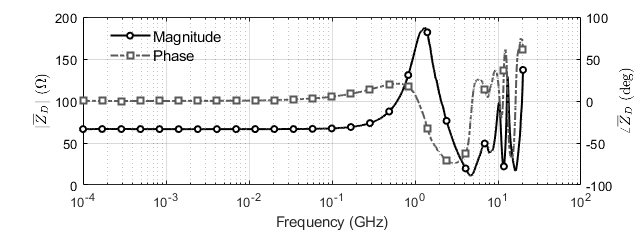}
    \caption{Equivalent impedance of the through-hole resistor obtained from the mean calibrated reflection coefficient measured with the MEMS-switched fixture.}
    \label{fig:zdut_mems}
\end{figure}

\begin{figure}[htbp]
    \centering
    \includegraphics[width=\columnwidth]{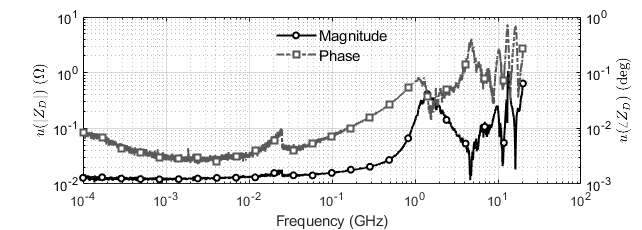}
    \caption{Type-A standard uncertainty of the equivalent DUT impedance obtained by propagating the Cartesian covariance matrix of the calibrated reflection coefficient through the impedance transformation.}
    \label{fig:zdut_uncertainty}
\end{figure}

The calibrated reflection coefficient was converted into the equivalent impedance referred to a 50~$\Omega$ line according to
\begin{equation}
    \cplx{Z}_{D}(f_k)
    =
    Z_0
    \frac{1+\overline{\cgamma}_{D}(f_k)}
    {1-\overline{\cgamma}_{D}(f_k)},
    \qquad Z_0=50~\Omega .
    \label{eq:gamma_to_impedance_results}
\end{equation}
The resulting impedance is reported in Fig.~\ref{fig:zdut_mems}. At low frequency, the impedance magnitude is close to the expected value for the mounted resistor. At higher frequencies, the response is dominated by the non-ideal behavior of the through-hole component and its connection to the fixture. The Type-A uncertainty associated with the impedance was obtained by propagating the Cartesian covariance matrix of the calibrated reflection coefficient through Eq.~\eqref{eq:gamma_to_impedance_results}. Figure~\ref{fig:zdut_uncertainty} reports the Type-A standard uncertainty of the impedance magnitude and phase.

\begin{figure}[tbp]
    \centering
    \includegraphics[width=\columnwidth]{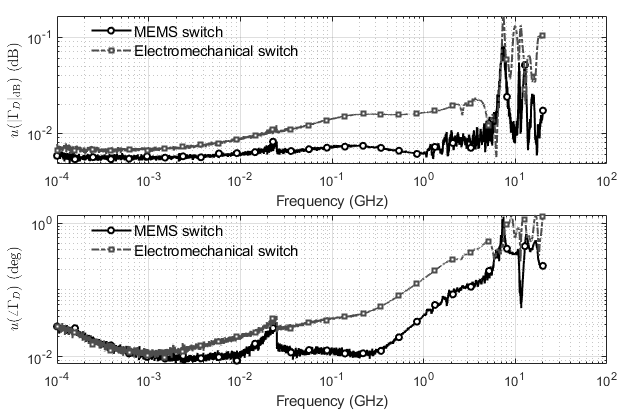}
    \caption{Scalar Type-A uncertainty indicators obtained by propagating the Cartesian covariance of the calibrated reflection coefficient to magnitude and phase.}
    \label{fig:unc_gamma_comparison}
\end{figure}

The repeatability of the switched calibration system is then assessed by comparison with the electromechanical switch. 
The relevant quantity for the present work is the dispersion of the calibrated complex reflection coefficient around its own mean value for each switching platform. 
For the electromechanical-switch measurements, the device was preconditioned by performing 50 preliminary commutations before starting the acquisition sequence, since a measurable temperature variation was observed during operation and could contribute to slow drifts of the RF response. 
For each frequency point, the covariance matrix $\mat{v}_{A}(f_k)$ is evaluated from the repeated calibrated values of $\hat{\cgamma}_{D}^{(i)}(f_k)$ in Cartesian coordinates. 
The corresponding covariance of the mean value, $\mat{U}_{A}(\overline{\cgamma}_{D})(f_k)$, is then obtained as described in Section~\ref{sec:metrological_model}. 
This covariance matrix is propagated to polar coordinates to obtain the Type-A standard uncertainties associated with the magnitude and phase of the mean calibrated DUT reflection coefficient. 
Figure~\ref{fig:unc_gamma_comparison} reports these uncertainties, namely $u(|\overline{\cgamma}_{D}|)$ and $u(\angle\overline{\cgamma}_{D})$, for the two switching implementations.
The MEMS-switched fixture exhibits a lower standard deviation over most of the investigated frequency range. At low frequency, the phase-related repeatability is comparable for the two systems. This is consistent with a regime in which the switching contribution is no longer dominant and the observed dispersion is mainly limited by the VNA trace noise. As frequency increases, the standard deviation grows for both systems. This increase is expected because phase errors produced by small variations of electrical length scale with frequency, while residual mismatch, connector parasitics, and switch-state dependent discontinuities generate stronger ripple and higher sensitivity in the corrected reflection coefficient.

\begin{figure}[htbp]
    \centering
    \includegraphics[width=\columnwidth]{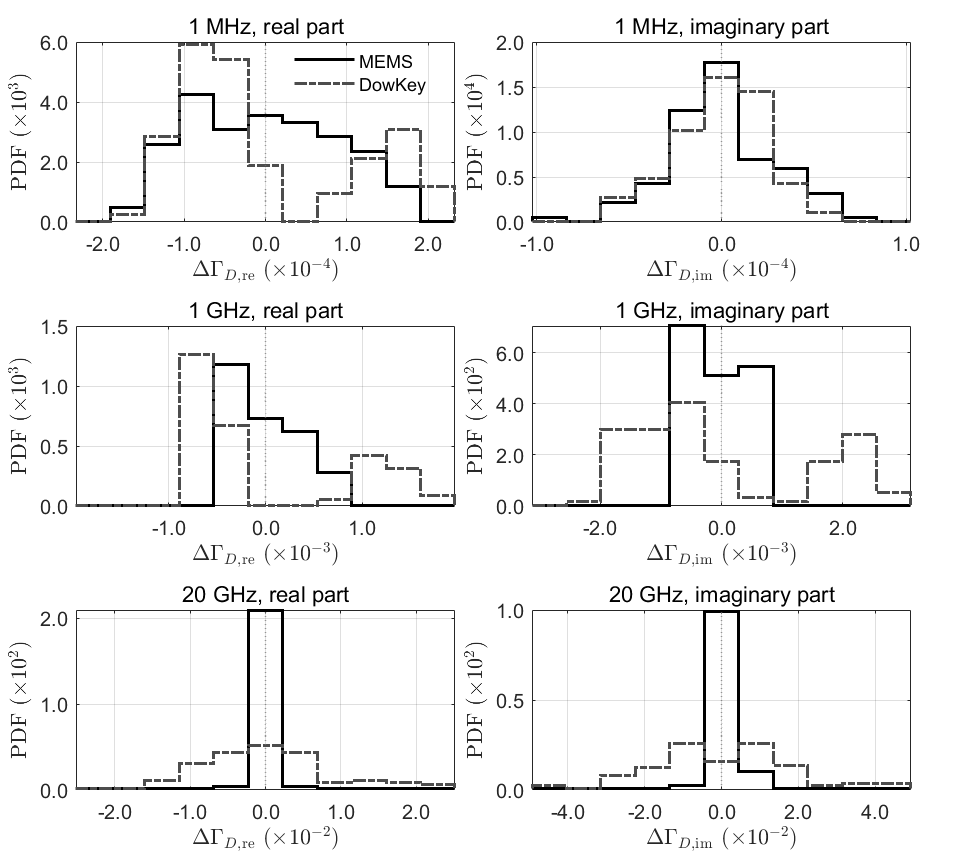}
    \caption{Probability density functions (PDFs) estimated from the centered real and imaginary components of the calibrated DUT reflection coefficient at 1~MHz, 1~GHz, and 20~GHz. The MEMS and electromechanical datasets are centered with respect to their own mean values before comparison.}
    \label{fig:histograms_comparison}
\end{figure}

A more direct view of the repeatability is provided by the histograms in Fig.~\ref{fig:histograms_comparison}, obtained from the centered quantities
$
    \Delta\cgamma_D^{(i)}(f_k)
    =
    \hat{\cgamma}_D^{(i)}(f_k)
    -
    \overline{\cgamma}_D(f_k)
$.
This removes the different mean DUT responses of the two fixtures and isolates the dispersion around each mean value. The selected frequencies, 1~MHz, 1~GHz, and 20~GHz, sample the low-frequency region, an intermediate microwave region, and the upper end of the investigated band. At low frequency, the two switching implementations provide comparable distributions. At higher frequencies, instead, the MEMS fixture exhibits more compact and nearly single-mode distributions, whereas the electromechanical switch shows broader, and in some cases bimodal, histograms. This suggests that the larger dispersion observed in the electromechanical case is not only due to Gaussian trace noise, but also to switching-state variability, contact repeatability, or mechanical settling effects.

\begin{figure}[htbp]
    \centering
    \includegraphics[width=\columnwidth]{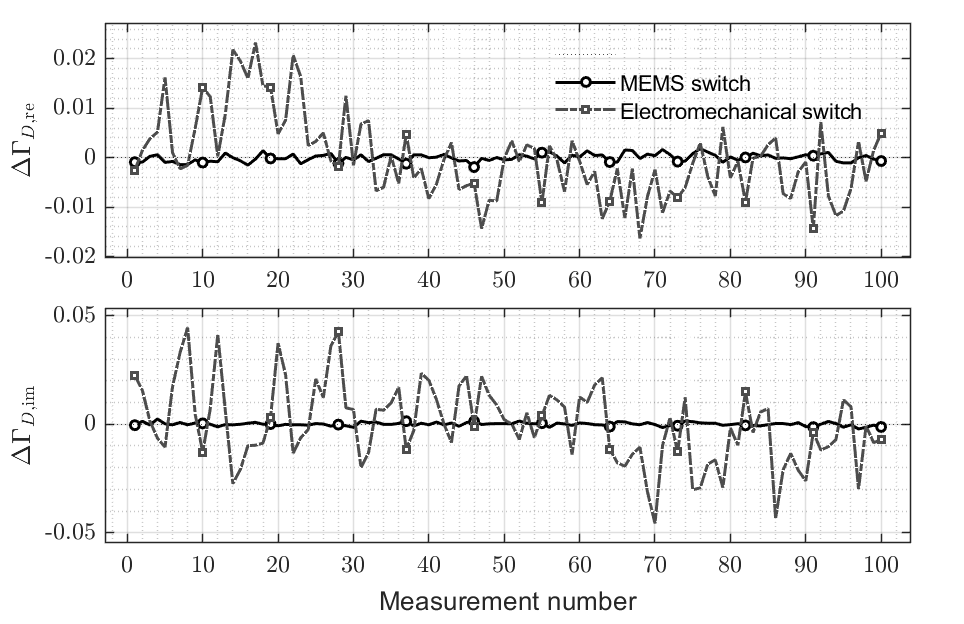}
    \caption{Centered real and imaginary components of the calibrated DUT reflection coefficient at 20~GHz as a function of the measurement number. The electromechanical switch was preconditioned by 50 preliminary commutations before the acquisition sequence.}
    \label{fig:drift_10ghz}
\end{figure}

The same conclusion is supported by the time evolution of the centered calibrated reflection coefficient. Figure~\ref{fig:drift_10ghz} reports the real and imaginary components of $\Delta\cgamma_D^{(i)}(f_k)$ at 20~GHz as a function of the measurement number. Despite the 50 preliminary commutations performed before acquisition, the electromechanical switch still exhibits long-term fluctuations, which means a reduced temporal stability. By contrast, the MEMS-switched fixture remains confined within a narrower interval over the full sequence of 100 measurements, with no comparable long-term drift.

\section{Conclusions}
\label{sec:conclusions}

The experimental results show that the MM5230-based RF MEMS switching platform provides repeatability at least comparable to, and over most of the investigated band better than, the Dow-Key electromechanical switch used as reference in this work. At low frequency, the improvement is not particularly evident: at 1~MHz, the standard deviation of the  $\Delta\mathit\Gamma_{D,\mathrm{re}}$ sample is $7.7\times10^{-5}$ for the MM5230 fixture and $1.1\times10^{-4}$ for the Dow-Key switch. However, the difference becomes much more pronounced at higher frequency. At 20~GHz, the corresponding values are $7.9\times10^{-4}$ and $8.4\times10^{-3}$, respectively, corresponding to more than one order of magnitude reduction for the MM5230-based fixture. This trend is also visible in Fig.~\ref{fig:unc_gamma_comparison}, where the repeatability curves obtained with the two switching implementations are close at low frequency and progressively separate as the frequency increases.

The centered histograms in Fig.~\ref{fig:histograms_comparison} further clarify this behavior. The MM5230-switched measurements are generally more compact and closer to single-mode distributions, whereas the Dow-Key data are broader and, in some cases, multimodal. This indicates that the Dow-Key implementation can introduce different effective switching states, which were not observed for the MM5230-based fixture within the repeated measurement sequence.

The improvement is therefore not limited to a reduction of random dispersion. The long-term stability analysis showed that the Dow-Key switch required a preliminary warm-up procedure before data acquisition, because a measurable temperature variation was observed during operation due to the large power dissipated by the driving coils (about 3 W in steady state). By contrast, thermal effects are expected to be negligible for the MM5230 under the present switching conditions. Considering the reported switching energy of 61~pJ per event and the switching rate used in the experiment, approximately 0.5~Hz, the average dynamic power associated with the RF MEMS actuation is only about 30~pW. This very low average dissipated power is particularly relevant for thermally constrained systems, where thermal stability of the calibration fixture is critical.

In view of these results, future work will focus on the trueness of the measurement by using calibrated reference loads and by including a complete uncertainty budget based on traceable SOL standard definitions.

\section*{Acknowledgment}
We gratefully acknowledge Tommaso Di Mascio for his kind and generous support to this work.

\bibliographystyle{IEEEtran} 
\bibliography{biblio}

@article{hall2016evaluating,
  title={Evaluating the measurement uncertainty of complex quantities: a selective review},
  author={Hall, BD},
  journal={Metrologia},
  volume={53},
  number={1},
  pages={S25--S31},
  year={2016},
  publisher={IOP Publishing}
}

@book{simons2001coplanar,
  author    = {Rainee N. Simons},
  title     = {Coplanar Waveguide Circuits, Components, and Systems},
  publisher = {John Wiley \& Sons},
  year      = {2001},
  chapter   = {3},
  pages     = {87--111}
}

@article{sain2016impact,
  author  = {A. Sain and K. L. Melde},
  title   = {Impact of Ground Via Placement in Grounded Coplanar Waveguide Interconnects},
  journal = {IEEE Transactions on Components, Packaging and Manufacturing Technology},
  volume  = {6},
  number  = {1},
  pages   = {136--144},
  year    = {2016}
}

@manual{MM101datasheet,
  title        = {{MM101 -- 8 Channel MEMS High Voltage Driver Datasheet}},
  organization = {Menlo Microsystems, Inc.},
  year         = {2025},
  month        = feb,
  note         = {Version 1.2, February 18, 2025}
}

@manual{MM5230datasheet,
  title        = {{MM5230 -- SP4T RF MEMS Switch Datasheet}},
  organization = {Menlo Microsystems, Inc.},
  year         = {2025},
  month        = mar,
  note         = {Version 2.0, March 19, 2025}
}

@Article{boretti2026relaymatrix,
AUTHOR = {Boretti, Edoardo and Torokhtii, Kostiantyn and Silva, Enrico and Alimenti, Andrea},
TITLE = {Scalable Relay Switching Platform for Automated Multi-Point Resistance Measurements},
JOURNAL = {Instruments},
VOLUME = {10},
YEAR = {2026},
NUMBER = {1},
ARTICLE-NUMBER = {3},
ISSN = {2410-390X}
}

@Article{shao2024rfmemssatellite,
AUTHOR = {Shao, Bingqian and Lu, Chengjian and Xiang, Yinjie and Li, Feixiong and Song, Mingxin},
TITLE = {Comprehensive Review of RF MEMS Switches in Satellite Communications},
JOURNAL = {Sensors},
VOLUME = {24},
YEAR = {2024},
NUMBER = {10},
ARTICLE-NUMBER = {3135},
PubMedID = {38793988},
ISSN = {1424-8220}
}

@article{spietz2026cryogenic,
  title={Cryogenic RF MEMS Switch With Electronic Calibration Capability},
  author={Spietz, Lafe F and Giovanniello, Chris and Takaki, Brandon and Ye, Wei and Boiko, Beverly and Long, Christian J and Flowers-Jacobs, Nathan E and Sirois, Adam J and Hopkins, Peter F and Benz, Samuel P and others},
  journal={IEEE Journal of Microwaves},
  volume={6},
  number={1},
  pages={251--262},
  year={2026},
  publisher={IEEE}
}

@inproceedings{brewer2025demonstrated,
  title={Demonstrated Reliability of the Ideal Switch{\textregistered}-Glass Based MEMS Device},
  author={Brewer, J and Nassar, C and Shorey, A and Gu, Y},
  booktitle={2025 IEEE International Reliability Physics Symposium (IRPS)},
  pages={1--5},
  year={2025},
  organization={IEEE}
}

@article{daneshmand2011rf,
  title={RF MEMS satellite switch matrices},
  author={Daneshmand, Mojgan and Mansour, Raafat R},
  journal={IEEE Microwave Magazine},
  volume={12},
  number={5},
  pages={92--109},
  year={2011},
  publisher={IEEE}
}

@article{bardin2021microwaves,
  title={Microwaves in quantum computing},
  author={Bardin, Joseph C and Slichter, Daniel H and Reilly, David J},
  journal={IEEE Journal of Microwaves},
  volume={1},
  number={1},
  pages={403--427},
  year={2021},
  publisher={IEEE}
}

@article{yeh2013situ,
  title={In situ broadband cryogenic calibration for two-port superconducting microwave resonators},
  author={Yeh, Jen-Hao and Anlage, Steven M},
  journal={Review of Scientific Instruments},
  volume={84},
  number={3},
  year={2013},
  publisher={AIP Publishing}
}

@ARTICLE{thomas2025cryogenic,
  author={Thomas, Jeremy N. and Hoffmann, Johannes and Flowers-Jacobs, Nathan E. and Fox, Anna E. and Jungwirth, Nicholas R. and Johnson-Wilke, Raegan L. and Dresselhaus, Paul D. and Benz, Samuel P.},
  journal={IEEE Transactions on Microwave Theory and Techniques}, 
  title={Cryogenic On-Chip In Situ S-Parameter Calibration Using Superconducting Coplanar Waveguides}, 
  year={2025},
  volume={73},
  number={11},
  pages={8942-8955},
  doi={10.1109/TMTT.2025.3585803}}

@article{hindle2010state,
  title={The state of {RF}/microwave switches},
  author={Hindle, Pat},
  journal={Microwave Journal},
  volume={53},
  number={11},
  pages={20--36},
  year={2010},
  publisher={Horizon House Publications, Inc., 685 Canton Street Norwood MA 02062 USA}
}

@article{cory2009rf,
  title={{RF}/microwave solid state switches},
  author={Cory, Rick and Fryklund, D},
  journal={MPD/Microwave Product Digest},
  year={2009}
}

@inproceedings{grant2004comparison,
  title={A comparison between {RF MEMS} switches and semiconductor switches},
  author={Grant, PD and Denhoff, MW and Mansour, Raafat R},
  booktitle={2004 International Conference on MEMS, NANO and Smart Systems (ICMENS'04)},
  pages={515--521},
  year={2004},
  organization={IEEE}
}

@article{kurmendra2024mems,
  title={{MEMS} switch realities: Addressing challenges and pioneering solutions},
  author={Kurmendra and Agarwal, Saurabh},
  journal={Micromachines},
  volume={15},
  number={5},
  pages={556},
  year={2024},
  publisher={MDPI}
}

@article{arakawa2023calibrated,
  title={Calibrated two-port microwave measurement up to {26.5 GHz for wide temperature range from 4 to 300 K}},
  author={Arakawa, Tomonori and Kon, Seitaro},
  journal={IEEE Transactions on Instrumentation and Measurement},
  volume={72},
  pages={1--8},
  year={2023},
  publisher={IEEE}
}

@article{firebaugh2004optical,
  title={Optical deflection measurement for characterization of microelectromechanical systems (MEMS)},
  author={Firebaugh, Samara L and Charles, HK and Edwards, Richard L and Keeney, Allen C and Wilderson, Samuel F},
  journal={IEEE Transactions on Instrumentation and Measurement},
  volume={53},
  number={4},
  pages={1047--1051},
  year={2004},
  publisher={IEEE}
}

@ARTICLE{madonna2002testing,
  author={Madonna, G. and Ferrero, A. and Pirola, M. and Pisani, U.},
  journal={IEEE Transactions on Instrumentation and Measurement}, 
  title={Testing microwave devices under different source impedance values-a novel technique for on-line measurement of source and device reflection coefficients}, 
  year={2000},
  volume={49},
  number={2},
  pages={285-289}}

@book{dunsmore2020handbook,
  title={Handbook of microwave component measurements: with advanced VNA techniques},
  author={Dunsmore, Joel P},
  year={2020},
  publisher={John Wiley \& Sons}
}

@article{lee2026cryogenic,
  title={Cryogenic performance evaluation of commercial {SP4T} microelectromechanical switch for quantum computing applications},
  author={Lee, Yong-Bok and Devitt, Connor and Zhu, Xu and Yost, Nicholas and Gu, Yabei and Bhave, Sunil A},
  journal={Microsystems \& Nanoengineering},
  volume={12},
  number={1},
  pages={72},
  year={2026},
  publisher={Nature Publishing Group UK London}
}

@article{spietz2025memsduino,
  title={MEMSDuino: An Arduino-Based MEMS Switch Controller},
  author={Spietz, Lafe and Sirois, Adam and Flowers-Jacobs, Nathan and Waltman, Steve and Benz, Samuel and Hopkins, Peter},
  journal={arXiv preprint arXiv:2501.03340},
  year={2025}
}

@article{vasjanov2024unveiling,
  title={Unveiling the sub-10 {GHz performance of SMA} connectors: A comparative analysis},
  author={Vasjanov, Aleksandr and Barzdenas, Vaidotas and Jurgo, Marijan and Gursnys, Darius},
  journal={Electronics},
  volume={13},
  number={14},
  pages={2686},
  year={2024},
  publisher={MDPI}
}

@article{zeier2018guidelines,
  title={Guidelines on the evaluation of vector network analysers ({VNA})},
  author={Zeier, Markus and Allal, Djamel and Judaschke, Rolf},
  journal={EURAMET Calibration Guide},
  volume={3},
  number={12},
  pages={507--521},
  year={2018}
}

@techreport{jcgm102_2011,
  title = {Evaluation of measurement data — Supplement 2 to the “Guide to the expression of uncertainty in measurement” — Extension to any number of output quantities},
  author = {{JCGM}},
  institution = {Joint Committee for Guides in Metrology (JCGM)},
  year = {2011},
  type = {JCGM},
  number = {102:2011},
  url = {https://www.bipm.org/en/committees/jc/jcgm/publications}
}

@manual{macomMA4AGSW4,
  title        = {{MA4AGSW4 SP4T AlGaAs PIN Diode Switch Datasheet}},
  organization = {MACOM Technology Solutions},
  note         = {Electrical specifications with bias current of the order of 10 mA},
  year         = {2023}
}

@INPROCEEDINGS{schudak2025glass,
  author={Schudak, Svenja and Bertke, Maik and Heinz, Jannis B.},
  booktitle={MikroSystemTechnik KONGRESS 2025; Mikroelektronik/Mikrosystemtechnik und ihre Anwendungen – Nachhaltigkeit und Technologiesouveränität}, 
  title={A glass-based {MEMS switch approach for harsh environments and high-performance RF }applications}, 
  year={2025},
  volume={},
  number={},
  pages={104-107},
  doi={10.30420/456614076}}

@article{kitano2008,
  title = {Broadband Method for Precise Microwave Spectroscopy of Superconducting Thin Films near the Critical Temperature.},
  author = {Kitano, Haruhisa and Ohashi, Takeyoshi and Maeda, Atsutaka},
  year = 2008,
  month = jul,
  journal = {Rev. Sci. Instrum.},
  volume = {79},
  number = {7},
  eprint = {18681723},
  eprinttype = {pubmed},
  pages = {074701},
  issn = {0034-6748},
  doi = {10.1063/1.2954957},
  pmid = {18681723}
}

@article{zinsser2019,
  title = {Role of Non-Linear Effects and Standing Waves in Microwave Spectroscopy: {{Corbino}} Measurements on Superconductors and {{VO2}}},
  shorttitle = {Role of Non-Linear Effects and Standing Waves in Microwave Spectroscopy},
  author = {Zin{\ss}er, Mario and Schlegel, Katrin and Dressel, Martin and Scheffler, Marc},
  year = 2019,
  month = mar,
  journal = {Review of Scientific Instruments},
  volume = {90},
  number = {3},
  pages = {034704},
  issn = {0034-6748},
  doi = {10.1063/1.5063862},
  urldate = {2026-07-03},
}

@article{silva2016,
  title = {Wideband {{Surface Impedance Measurements}} in {{Superconducting Films}}},
  author = {Silva, Enrico and Pompeo, Nicola and Torokhtii, Kostiantyn and Sarti, Stefano},
  year = 2016,
  journal = {IEEE Trans. Instrum. Meas.},
  volume = {65},
  number = {5},
  pages = {1120--1129}
}

@article{booth1994,
  title = {A Broadband Method for the Measurement of the Surface Impedance of Thin Films at Microwave Frequencies},
  author = {Booth, J C and Wu, Dong Ho and Anlage, Steven M},
  year = 1994,
  journal = {Rev. Sci. Instrum.},
  volume = {65},
  number = {6},
  pages = {2082--2090}
}

\end{document}